\pdfoutput=1
\documentclass[conference]{IEEEtran}
\IEEEoverridecommandlockouts

\usepackage{graphicx}
\usepackage{textcomp}
\def\BibTeX{{\rm B\kern-.05em{\sc i\kern-.025em b}\kern-.08em
    T\kern-.1667em\lower.7ex\hbox{E}\kern-.125emX}}

\usepackage{cite}
\usepackage[utf8]{inputenc} 
\usepackage[T1]{fontenc}    
\usepackage{hyperref}       
\usepackage{url}            
\usepackage{booktabs}       
\usepackage{amsfonts}       
\usepackage{nicefrac}       
\usepackage{microtype}      
\usepackage{subcaption}
\usepackage{ifthen}
\usepackage{bbm}
\usepackage{amssymb}
\usepackage{algorithmicx}
\usepackage[cmex10]{amsmath}
\usepackage{amsthm}
\usepackage{mathtools}
\usepackage{bm}
\usepackage[inline, shortlabels]{enumitem}
\usepackage[electronic]{ifsym}
\usepackage{physics}

\usepackage{makecell,pbox,ragged2e,hhline}
\usepackage[table, x11names, svgnames]{xcolor}
\usepackage{color,diagbox,tabularx}

\usepackage{tikz}
\usetikzlibrary{fit, shapes.geometric, shapes.misc, arrows.meta, positioning, decorations.markings, matrix}
\tikzset{>={Latex[width=0.5mm,length=1.2mm]}}
\usepackage{smartdiagram}
\usesmartdiagramlibrary{additions}
\usepackage{pgfplots}
\pgfplotsset{compat=newest, ticks=none}
\usepackage{varwidth}
\tikzset{every picture/.style={/utils/exec={\fontfamily{lmss}}}}

\usepackage{steinmetz}
\usepackage{xpatch}
\xpatchcmd{\phase}{#2}{\hspace{0.8pt}\vphantom{\scalebox{0.8}{\tiny{,}}}#2\hspace{1.4pt}}{}{}

\makeatletter
\newdimen\@widthOfTo%
\newdimen\@widthOfImplies%
\pgfmathsetmacro{\@scaleFactorImplies}{\@widthOfTo/\@widthOfImplies}%
\newcommand*{\ScaledImplies}{\mathrel{\raisebox{0.3ex}{\scalebox{\@scaleFactorImplies}{\ensuremath{\Longrightarrow}}}}}%
\makeatother

\newcolumntype{x}[1]{>{\centering\arraybackslash}p{#1}}

\usepackage{colortbl}
\definecolor{LightCyan}{rgb}{0.7,1,1}
\definecolor{LightGrey}{rgb}{0.95,0.95,0.95}
\definecolor{DarkBlue}{rgb}{0,0,0.4}
\definecolor{Yellow}{rgb}{1,1,0.6}

\DeclareMathOperator{\modrelu}{ModReLU}
\DeclareMathOperator{\crelu}{CReLU}
\DeclareMathOperator{\re}{Re}
\DeclareMathOperator{\im}{Im}

\let\originalleft\left
\let\originalright\right
\renewcommand{\left}{\mathopen{}\mathclose\bgroup\originalleft}
\renewcommand{\right}{\aftergroup\egroup\originalright}

\theoremstyle{plain}

\theoremstyle{definition}

\theoremstyle{remark}

\usepackage{listings}
\lstdefinestyle{python}{
  belowcaptionskip=1\baselineskip,
  breaklines=true,
  frame=shadowbox,
  rulesepcolor=\color{gray},
  xleftmargin=\parindent,
  language=Python,
  showstringspaces=false,
  basicstyle=\footnotesize\ttfamily,
  keywordstyle=\bfseries\color{deepblue},
  moredelim=**[s][\color{blue}]{'''}{'''},
  commentstyle=\itshape\color{magenta},
  identifierstyle=\color{black},
  stringstyle=\color{red}
}
\lstdefinestyle{output}{
  belowcaptionskip=1\baselineskip,
  breaklines=true,
  frame=L,
  basicstyle=\footnotesize\ttfamily,
  xleftmargin=\parindent
}

\def\BibTeX{{\rm B\kern-.05em{\sc i\kern-.025em b}\kern-.08em T\kern-.1667em\lower.7ex\hbox{E}\kern-.125emX}}

\begin{document}

\title{Robust Wireless Fingerprinting via
\\  Complex-Valued Neural Networks
}
\author{
  \IEEEauthorblockN{Soorya Gopalakrishnan\IEEEauthorrefmark{1}\thanks{\IEEEauthorrefmark{1}Joint first authors.}, Metehan Cekic\IEEEauthorrefmark{1}, Upamanyu Madhow}

  \IEEEauthorblockA{\textit{Department of Electrical and Computer Engineering}\\
  					\textit{University of California, Santa Barbara}\\
                    Email: \{soorya, metehancekic, madhow\}@ucsb.edu}
}

\maketitle


\begin{abstract}
A ``wireless fingerprint'' which exploits hardware imperfections unique to each device is a potentially powerful tool for wireless security.
Such a fingerprint should be able to distinguish between devices sending the same message, and should be robust against standard
spoofing techniques. Since the information in wireless signals resides in complex baseband, in this paper, we explore the use of neural 
networks with complex-valued weights to learn fingerprints using supervised learning.  We demonstrate that, 
while there are potential benefits to using sections of the signal beyond just the preamble to learn fingerprints, 
the network cheats when it can, using information such as transmitter ID (which can be easily spoofed) to artificially inflate performance. 
We also show that noise augmentation by inserting additional white Gaussian noise can lead to significant performance gains, which indicates that
this counter-intuitive strategy helps in learning more robust fingerprints. We provide results for two different wireless protocols, 
WiFi and ADS-B, demonstrating the effectiveness of the proposed method.
\end{abstract}



\section{Introduction} \label{sec:intro}

With the proliferation of wireless devices in everyday life, assuring the security of such devices becomes a critical concern.
We focus here on a potentially powerful tool for this purpose: wireless fingerprints based on hardware imperfections unique to each device.
Prior work shows that it is possible to extract such fingerprints, but it is often based on handcrafted features 
extracted with knowledge of the underlying protocol \cite{brik2008wireless,hua2018accurate}. In this paper, we investigate the use of a protocol-agnostic
approach, employing supervised learning of fingerprints via a neural network.

Our goal is to extract a fingerprint that enables us to distinguish between two devices sending
exactly the same message, using as input the complex baseband signal at the receiver.  
Since the input is complex-valued and one-dimensional (1D), we employ a 1D convolutional neural network (CNN) with complex-valued weights.
When compared to prior approaches \cite{o2016convolutional,sankhe2018oracle} that use real-valued networks (with real and imaginary parts of input data treated as independent channels), these networks have a smaller degree of freedom available at the synaptic level. 
It has been observed that this confers generalization benefits \cite{hirose2012generalization}.

While we would like to develop wireless fingerprinting  techniques that are protocol-agnostic, we must remain vigilant against locking onto easily spoofed features.
A naive protocol-agnostic scheme would not distinguish between any segments of the message from which the fingerprint is being extracted.
However, for any communication protocol, the message contains transmitter ID information, e.g.\  the MAC address in WiFi packets, the ICAO aircraft address in ADS-B (Automatic Dependent Surveillance-Broadcast) air traffic control signals, etc. Such ID information can be spoofed, hence any fingerprinting technique that uses the entire message must prove that it does not ``cheat'' by focusing just on the ID.  
We demonstrate that a completely protocol-agnostic CNN is vulnerable to such involuntary cheating, and then show that
using the preamble, which is common to all packets from all transmitters, suffices to obtain reasonable accuracies, despite the relatively short length of the preamble
compared to the length of the entire message. We then explore the impact of noise on training, and propose a noise augmentation strategy for enhancing performance.


\begin{figure*}[]
\centering
\begin{minipage}{0.92\textwidth}
\begin{tikzpicture}[every node/.style={inner sep=0,outer sep=1.5pt}]
	\node[anchor=south west,inner sep=0] (image) at (0,0) {\includegraphics[width=0.9\textwidth]{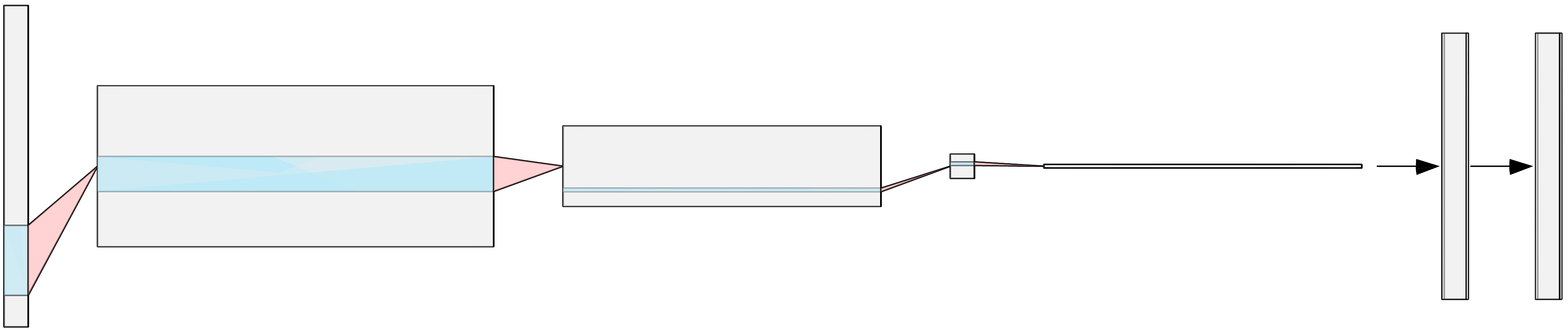}};
	\begin{scope}[x={(image.south east)},y={(image.north west)}]
			\path (-0.009, 0.02) -- (-0.009, 0.97) node[midway, color=black!70] (ip) {\scalebox{1.0}{\tiny{320}}};
				\draw [<-,  color=black!70]  (-0.009, 0.02) --(ip.south);
				\draw [->,  color=black!70] (ip.north) -- (-0.009, 0.97);
			\path (0.065, 0.685) -- (0.315, 0.685) node[midway, color=black!70] (c1w) {\scalebox{1.0}{\tiny{100}}};
				\draw [<-,  color=black!70]  (0.065, 0.685) --(c1w.west);
				\draw [->,  color=black!70] (c1w.east) -- (0.315, 0.685);
			\path (0.36, 0.57) -- (0.56, 0.57) node[midway, color=black!70] (c2w) {\scalebox{1.0}{\tiny{100}}};
				\draw [<-,  color=black!70]  (0.36, 0.57) --(c2w.west);
				\draw [->,  color=black!70] (c2w.east) -- (0.56, 0.57);
			\filldraw[fill=gray!10, very thin] (0.663, 0.48) rectangle (0.868, 0.52);
	        \path (0.665, 0.57) -- (0.865, 0.57) node[midway, color=black!70] (avg) {\scalebox{1.0}{\tiny{100}}};
		        \draw [<-,  color=black!70]  (0.665, 0.57) --(avg.west);
		        \draw [->,  color=black!70] (avg.east) -- (0.865, 0.57);
		\node at (0, -0.25) (a) {\small 
			\begin{tabular}{l} 
				ADS-B preamble\\[3pt]
				\scriptsize{\begin{varwidth}{\linewidth} 
				\begin{itemize}[leftmargin=12pt, itemsep=1.5pt, label={--}] 
					\item 320 samples 
					\item Sampled at 20 MHz 
				\end{itemize}\end{varwidth}}
			\end{tabular}};
		\node at (0.195, -0.02) (a) {\small 
			\begin{tabular}{l} 
				Complex conv.\ layer \\[3pt]
				\scriptsize{\begin{varwidth}{\linewidth} 
				\begin{itemize}[leftmargin=20pt, itemsep=1.5pt, label={--}] 
					\item Kernel size: 40 
					\item Stride: 20
					\item  100 filters
				\end{itemize}\end{varwidth}}
			\end{tabular}};          
		\node at (0.46, 0.08) (a) {\small 
			\begin{tabular}{l}  
				Complex conv.\ layer \\[3pt] 
				\scriptsize{\begin{varwidth}{\linewidth} 
				\begin{itemize}[leftmargin=20pt, itemsep=1.5pt, label={--}] 
					\item Kernel size: 5 
					\item Stride: 1
					\item  100 filters
				\end{itemize}\end{varwidth}}
			\end{tabular}};    
		\node [draw, circle, fill=blue!10, minimum size=0.4cm, very thin] at (0.613, 0.5) {};
		\node at (0.62, 0.28) (a) {\small $|\cdot|^2$};
		\node at (0.77, 0.34) (a) {\small Temporal average};
		\node at (0.878, -0.14) (a) {\small 
			\begin{tabular}{l} 
				Real dense \\layer \\[0pt] 
				\scriptsize{\begin{varwidth}{\linewidth} 
				\begin{itemize}[leftmargin=10pt, itemsep=1.5pt, label={--}] 
					\item 100 neurons
				\end{itemize}\end{varwidth}}
			\end{tabular}};
		\node at (0.98, -0.1) (a) {\small 
			\begin{tabular}{l} 
				Output \\ layer 
			\end{tabular}};
	\end{scope}
\end{tikzpicture}
\end{minipage}
\caption{Complex-valued 1D CNN architecture for ADS-B signals.}
\label{fig:arch}
\vspace{-2pt}
\end{figure*}
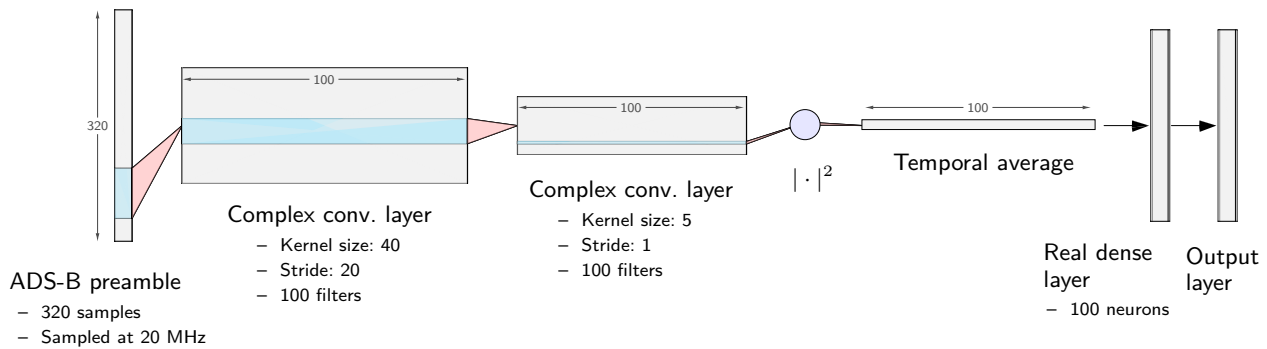

\subsection*{Contributions}

We propose a protocol-agnostic fingerprinting technique using complex-valued CNNs and demonstrate its robustness to various real-world imperfections. Our main contributions are as follows:

\begin{itemize}[itemsep=2pt, topsep=2pt, leftmargin=15pt]
\item We demonstrate that supervised learning using complex-valued CNNs works well for two different wireless protocols, WiFi and ADS-B, and compare the performance of different complex activation functions and architectures.

\item When making use of portions of the signal beyond just the preamble, we show that networks will ``cheat'' whenever given the chance, resulting in artificially high accuracies (that are independent of the noise level) by focusing on the transmitter ID information present in these sections. 

\item We then focus on learning fingerprints from the preamble.
Restricting to the preamble is not strictly protocol-agnostic, but, in principle, the location and extent of the preamble can be identified in unsupervised fashion for any given protocol by correlating packets across different transmitters.
We study the robustness of our approach to noise in the data, and find that performance is better when the training set has lower SNR than the test set.

\item We show that noise augmentation, or insertion of additional white Gaussian noise (AWGN), can significantly improve performance, presumably because it aids in learning more robust fingerprints. In particular, it is important to add noise to test data as well as the training data (with more noise added to training data) to yield benefits.

\end{itemize}


\section{Related Work}

Wireless device fingerprinting can be accomplished using either the transient (microsecond-length) signals transmitted during the on/off operation of wireless devices, or via the steady-state packet information present in between the start and end transients. We focus here on work that employs the steady-state method since it is of more practical utility \cite{kennedy2008radio}. Work in this area can be broadly divided into two categories: traditional approaches that use handcrafted features as device characteristics, and techniques that employ machine learning to obtain fingerprints.

\subsection{Traditional approaches}

Remote physical device fingerprinting using small, microscopic deviations in device hardware called clock skews was introduced in \cite{kohno2005remote}. The clock skew of a single device was observed to be fairly consistent over time, but clock skews varied significantly across devices, enabling fingerprinting. For wired devices in wide area networks, \cite{kohno2005remote} estimated clock skews using TCP/IP packet headers. This technique was extended by \cite{jana2010fast} to wireless local area networks where more accurate measurements are possible from the Time Synchronization Function timestamps in IEEE 802.11 frames. However, these two detection methods were defeated by \cite{arackaparambil2010reliability} which devised attacks to spoof the clock skew of a fake device to mimic that of a real one. Using more parameters such as jitter and fitting errors to measure the authenticity of the skew can mitigate these spoofing attempts. More recently, \cite{hua2018accurate} proposed using the carrier frequency offset (CFO) as a long-term device fingerprint, with the offset estimated using channel state information (CSI) measurements. While application layer spoofing of CFO is difficult \cite{hua2018accurate}, using the CFO as a mechanism for physical security has two key drawbacks: first, it does not provide a stable signature, since the oscillator frequency drifts over time; second, an adversary manipulating baseband signals can easily alter the CFO.
 
\subsection{Machine learning based approaches}

The first use of discriminatory classifiers for fingerprinting was in \cite{kennedy2008radio}, which used a $k$-nearest neighbor ($k$-NN) classifier after preprocessing WiFi data to extract the log-spectral-energy of the preamble. A different preprocessing step was proposed in \cite{brik2008wireless}, involving demodulation error metrics such as frequency offset and I/Q offset, followed by a support vector machine (SVM). For the ADS-B air traffic control protocol, \cite{strohmeier2015passive} performed $k$-means clustering on features based on inter-arrival times of aircraft position, velocity and identification messages. A similar inter-arrival approach was shown in \cite{radhakrishnan2015gtid} to be effective for WiFi fingerprinting, with a real-valued neural network (NN) operating on the extracted features. In \cite{leonardi2017air}, the carrier phase offset of ADS-B signals was used as input to an NN to learn fingerprints. 
For IEEE 802.15.4 ZigBee devices, \cite{merchant2018deep} proposed the use of a real-valued CNN operating on an error signal obtained by subtracting out the ideal estimated signal from received data. 
 These techniques work well, but they rely on protocol-specific signal modeling and preprocessing prior to learning, in contrast to our approach.
 
A purely learning based approach was studied in \cite{o2016convolutional,o2017introduction}, albeit for modulation recognition and not device fingerprinting. Each packet was sliced into multiple training examples using sliding windows, with the real and imaginary parts of complex data treated as independent channels. These were then input to a real-valued CNN capable of recognizing different analog and digital modulation types. The use of a real-valued CNN for WiFi device fingerprinting was studied in \cite{sankhe2018oracle}, with sliding window preprocessing similar to prior work. 
As discussed in Section \ref{sec:intro}, our proposed method of learning complex-valued representations has potential generalization benefits over real-valued approaches \cite{hirose2012generalization}.


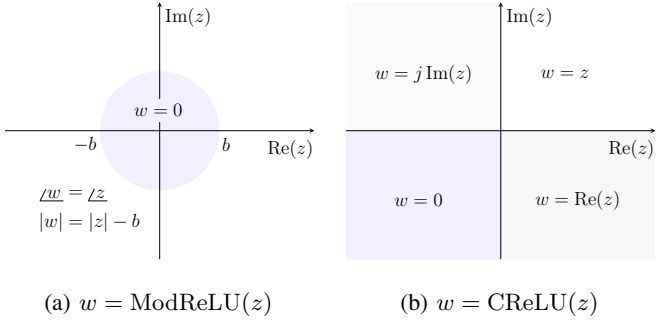
\begin{figure}[t]
\centering
\hspace{-5pt}
\begin{subfigure}{0.49\linewidth}
	\begin{center}
	\scalebox{0.6}{
	\begin{tikzpicture}
		\begin{axis}[
			axis lines = center,
			set layers=standard,
			xmin=-0.5, xmax=0.5, ymin=-0.5, ymax=0.5,
			axis equal,
			xlabel = {\large $\re(z)$},
			ylabel = {\large $\im(z)$},
			x label style= {at ={(axis cs: 0.62, -0.13)}},
			yticklabels={,,},
			]
			\begin{pgfonlayer}{axis background}
				\filldraw[fill=white!50, color=blue!6] (axis cs: 0, 0) circle [radius=0.23];
			\end{pgfonlayer}
			\node [fill=blue!6] at (axis cs: -0.005, 0.08) (a) {\large $w=0$};
			\node at (axis cs: 0.258, -0.053) (a) {\large $b$};
			\node at (axis cs: -0.29, -0.053) (a) {\large $-b$};
			\node at (axis cs: -0.28, -0.3) (a) {\large 
				$\begin{aligned} 
					\phase{w} &= \phase{z}    \\[2pt]
					|w| &= |z| - b
				\end{aligned}$
				};
		\end{axis}
	\end{tikzpicture}}
	\end{center}
	\caption{$w=\modrelu(z)$}
	\label{fig:modrelu}
\end{subfigure}
\hfill
\begin{subfigure}{0.49\linewidth}
	\begin{center}
	\scalebox{0.6}{
	\begin{tikzpicture}
	\begin{axis}[
			axis lines = center,
			set layers=standard,
			xmin=-0.5, xmax=0.5, ymin=-0.5, ymax=0.5,
			axis equal,
			xlabel = {\large $\re(z)$},
			ylabel = {\large $\im(z)$},
			x label style= {at ={(axis cs: 0.62, -0.13)}},
			yticklabels={,,},
			legend style={at={(-0.2, 0.2)}},
			]
			\begin{pgfonlayer}{axis background}
				\filldraw[fill=white!100, color=blue!6]
								(axis cs: -0.6, -0.5) rectangle (axis cs: 0, 0);
				\filldraw[fill=gray!4, draw=none]
								(axis cs: -0.6, 0) rectangle (axis cs:0 , 0.5);
				\filldraw[fill=gray!6, draw=none]
							(axis cs: 0, -0.5) rectangle (axis cs: 0.6, 0);
			\end{pgfonlayer}
		 	\node at (axis cs: -0.32, -0.27) (a) {\large 
		 		$w=0$
		 		};
			\node at (axis cs: -0.3, 0.22) (a) {\large 
				$w =  j \im(z)$
				};
			\node at (axis cs: 0.3, -0.27) (a) {\large 
				$w = \re(z)$		
				};
			\node at (axis cs: 0.25, 0.22) (a) {\large 
				$w=z$			
				};
	\end{axis} 
	\end{tikzpicture}}
	\end{center}
	\caption{$w=\crelu(z)$}
	\label{fig:crelu}
\end{subfigure}
\hspace{2pt}
\caption{ModReLU and CReLU activation functions in the complex plane. ModReLU preserves the phase of all inputs outside a disc of radius $b$, while CReLU distorts all phases outside $[0, \pi/2]$ (the first quadrant). Figure adapted from \cite{trabelsi2017deep}.}
\label{fig:activation_functions}
\end{figure}


\section{Complex Valued CNN Architecture}

\subsection{Overview}

We use neural networks with complex-valued weights and biases to learn features from complex-valued wireless signals. Such complex-valued embeddings have found use in speech, music and vision tasks \cite{wisdom2016full,trabelsi2017deep}.
Here we employ the framework of \cite{trabelsi2017deep} which performs complex backpropagation by using partial derivatives of the cost with respect to the real and imaginary parts of each parameter. 
We make use of 1D complex convolutional layers with the following choices of activation functions (depicted in Fig. \ref{fig:activation_functions}):
\begin{itemize}[itemsep=2pt, topsep=3pt, leftmargin=20pt]
\item \textit{ModReLU} - This function preserves input phase and affects only the absolute value. Here $b$ is a learned bias. 
\begin{equation*}
\modrelu(z) = \max\left(|z| - b, 0\right) \,\,e^{j\phase{z}}.
\end{equation*}
\item \textit{CReLU} - Unlike ModReLU this function does not preserve phase, with separate ReLUs applied on the real and imaginary parts of the input. The phase of the output is therefore limited to $[0, \pi/2]$.
\begin{equation*}
\crelu(z) = \max\left(\re(z), 0\right) + j\max\left(\im(z), 0\right).
\end{equation*}
The loss in phase information can be potentially compensated by using filters with a larger number of channels that are capable of providing phase derotation. 

\end{itemize}

\noindent Fig.\ \ref{fig:arch} depicts a sample complex convolutional architecture for ADS-B signals.
We use a series of complex 1D convolutions followed by an $|\cdot|^2$ layer to convert complex representations to real ones, and then a series of real-valued layers after a temporal averaging layer to obtain the fingerprint. 

\subsection{Performance} \label{sec:arch_perf}

We provide results for an external database for two different wireless protocols: 
WiFi 802.11a (5.8 GHz) and 802.11g (2.4 GHz) commercial off-the-shelf emitters with a signal bandwidth of 20 MHz, and ADS-B (1.09 GHz) narrowband air traffic control signals. 
We start by using only the preamble for fingerprinting, with signals normalized to unit power. When sampled at 20 MHz, the length of the preamble is 320 I/Q samples for both protocol types. 

We report accuracies for the following networks:
\begin{itemize}[leftmargin=10pt, itemsep=2pt, topsep=2pt]
\item \textit{ADS-B}: 100$\, C\,$40$\times$20 -- 100$\, C\,$5$\times$1 -- $|\cdot|^2$ -- Avg -- 100$\, D$.
\item \textit{WiFi}: 100$\, C\,$20$\times$10 -- 100$\, C\,$10$\times$1 -- $|\cdot|^2$ -- Avg -- 100$\, D$.
\end{itemize} 
\noindent The notation should be read as follows: 
$<$number of filters$>$$\, C\,$$<$convolution size$>$$\times$$<$stride$>$, and $<$number of neurons$>$$\,D$, where $C$ represents a convolutional layer and $D$ a fully connected layer, with  complex-valued layers prior to the $|\cdot|^2$ layer and real-valued layers afterward. `Avg' denotes a temporal averaging layer.
We train networks for 200 epochs with a batch size of 100, using the Adam optimizer with default hyperparameters and $\ell_2$ regularization constant of 10$^{\text{-3}}$. 

\begin{figure}[t]
\centering
\includegraphics[width=0.95\columnwidth]{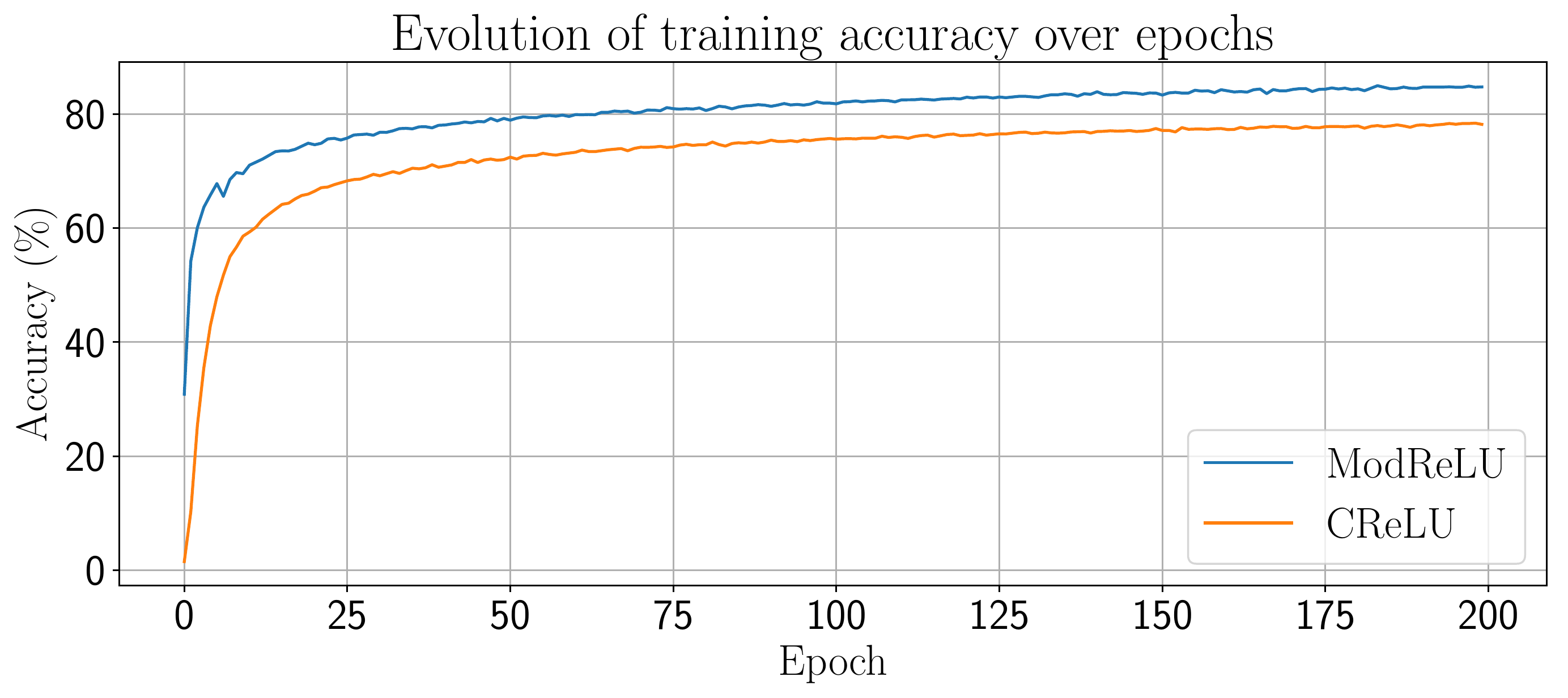}
\caption{Evolution of training accuracy over epochs for ModReLU and CReLU architectures (ADS-B, 100 devices). ModReLU provides a small (5\%) gain in train and test accuracies over CReLU, with similar convergence behavior.}
\label{fig_train_test_acc}
\end{figure}

We achieve 99.53\% fingerprinting accuracy for 19 WiFi devices without channel distortion, using 200 samples per device for training and 100 for testing. For the ADS-B protocol, we obtain 81.66\% accuracy with 100 devices (using 400 samples per device for training and testing).  Fig. \ref{fig_train_test_acc} compares the convergence of ModReLU and CReLU architectures. Both activation functions have similar convergence time, with ModReLU resulting in slightly higher accuracy for both the training and test sets.

\begin{table}[b]
\centering
\caption{Performance comparison between networks with complex and real weights (when using only the preamble).}
\label{table_complex_reim}
\begin{center}
\begin{small}
\begin{tabular}{llcr}
\toprule
Dataset & Network type & Accuracy & 
\begin{tabular}{@{}r@{}} No. of real \\ parameters \end{tabular} 
\\
\midrule
ADS-B & Complex & \textbf{81.66} & 128,400\\
& Real & 73.84 & 78,400\\
& Real (1.4x) & 73.25 & 133,680\\
& Real (2x) & 75.00 & 246,600\\
\midrule
WiFi & Complex & \textbf{99.53} & 216,219\\
& Real & 97.32 & 116,219\\
& Real (1.4x) & 97.53 & 217,899\\
& Real (2x) & 97.89 & 430,419\\
\bottomrule
\end{tabular}
\end{small}
\end{center}
\end{table}

\begin{figure*}[!t]
\centering
\hfill
\begin{subfigure}{0.265\linewidth}
	\includegraphics[width=0.46\linewidth]{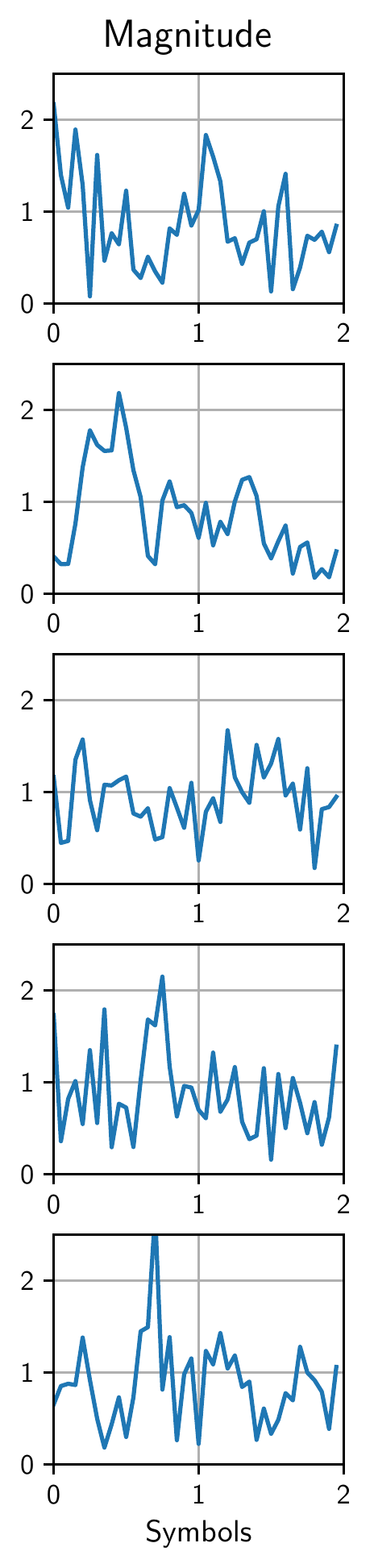}
	\includegraphics[width=0.46\linewidth]{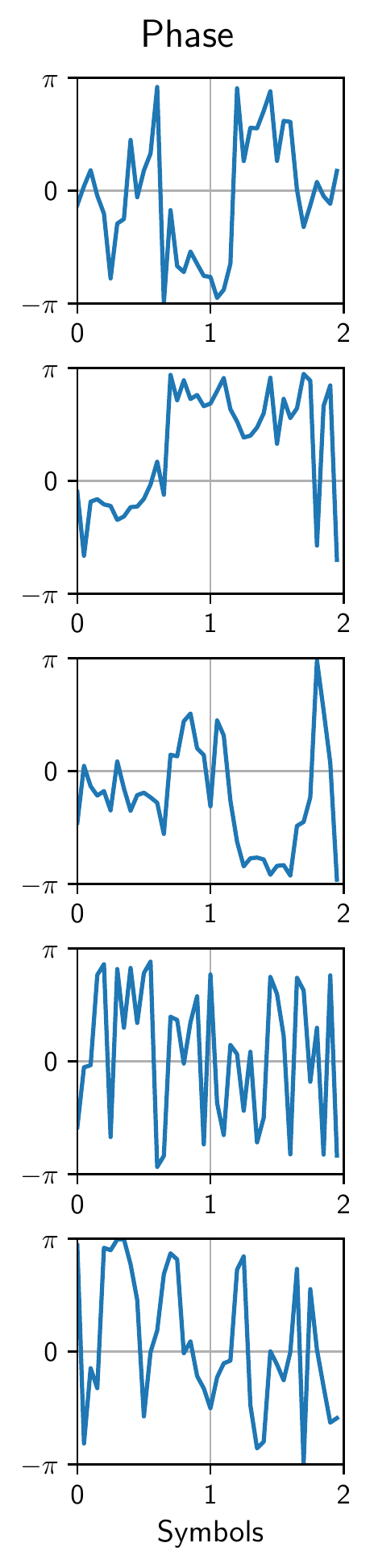}
 	\caption{Layer 1}
 	\label{fig:layer1} 
\end{subfigure}
\hspace{10pt}
\begin{subfigure}{0.68\linewidth}
	\includegraphics[width=0.46\linewidth]{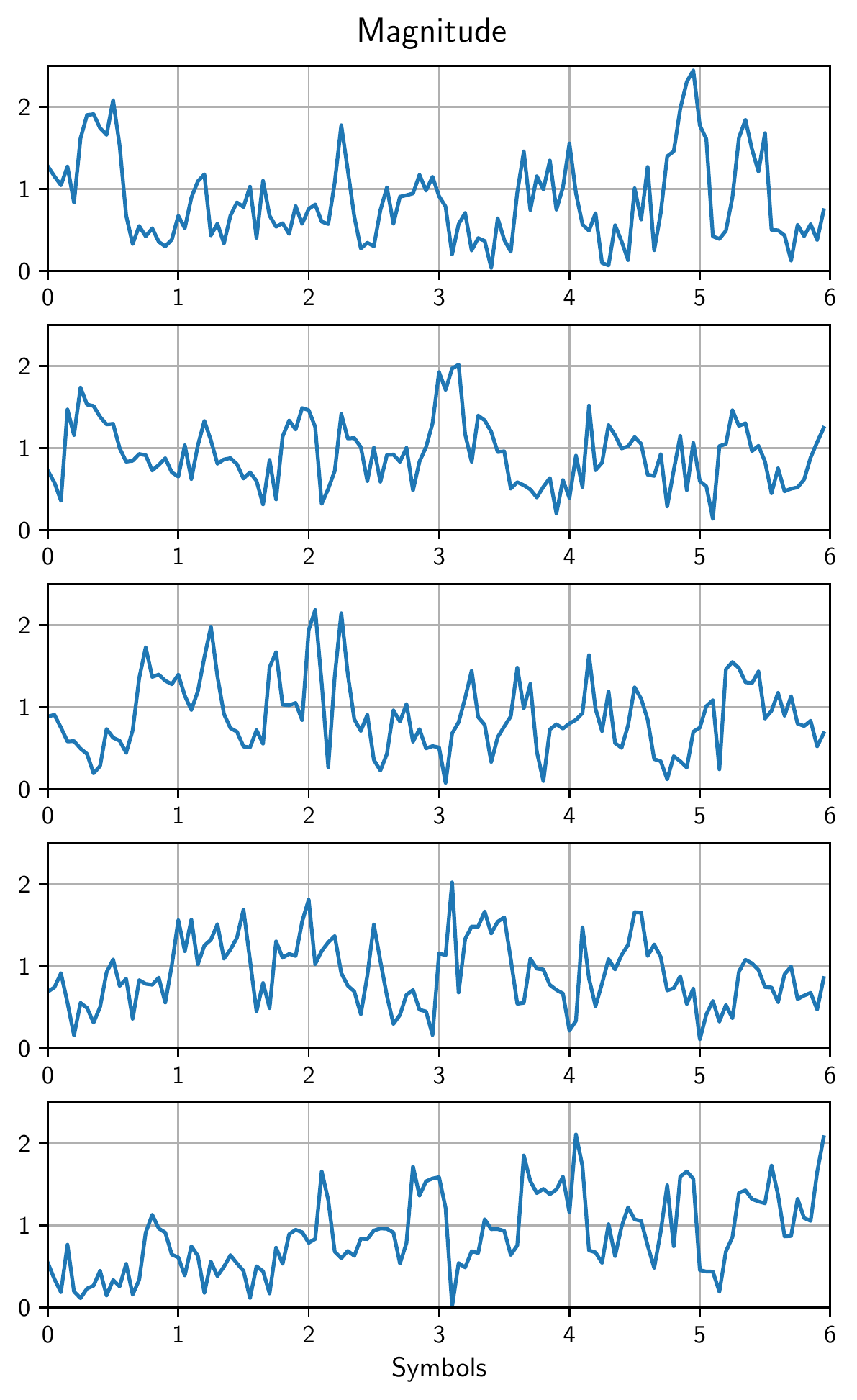}
	\includegraphics[width=0.46\linewidth]{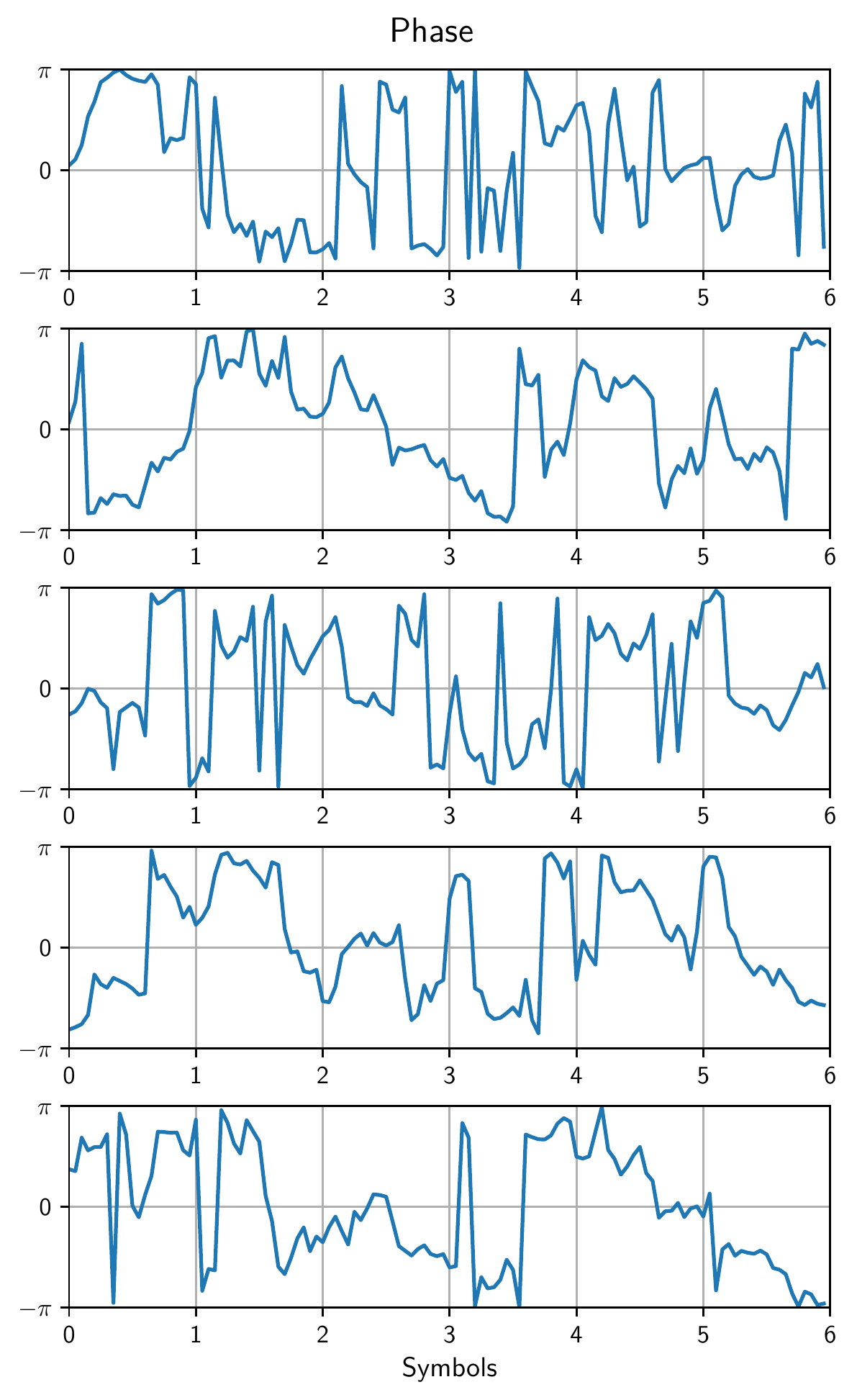}
	\caption{Layer 2}
	\label{fig:layer1} 
\end{subfigure}
\hfill
\caption{Visualizations of the first and second convolutional layer for ADS-B (ModReLU architecture). Each row shows the input signal that maximizes the activation of a particular filter, computed using gradient ascent starting from random noise. 
Convolutional filters in the first layer span 2 input symbols; filters in the second layer span 6 symbols.
}
\label{fig:visualize_layer_2}
\vspace{-5pt}
\end{figure*}

Table \ref{table_complex_reim} compares the performance of complex-valued and real-valued networks (for which real and imaginary parts of data are treated as different channels). If we fix the number of feature maps, a complex filter would contain twice as many parameters as an equivalent real filter. Therefore, we consider real networks where the number of channels is scaled by factors of 1, 1.4 and 2. We find that the complex network outperforms its real counterparts, with a gain in accuracy of 6.66\% for ADS-B and 1.64\% for WiFi.

Fig. \ref{fig:visualize_layer_2} visualizes the first and second convolutional layer of the ADS-B architecture, showing the input signal that maximizes the activations of each filter. 
Since transmitter-characteristic nonlinear effects manifest themselves primarily in short-term transitions of amplitude and phase, the filters in the first layer can capture these effects by spanning a small multiple of the symbol interval (2 symbols).
To compute these signals, we start from randomly generated noise and use 200 steps of gradient ascent to maximize the absolute value of each filter output, with the signal normalized to unit power at each step.


\section{Robustness to ID} \label{sec:id}

In this section, we investigate the potential benefits to using post-preamble portions of the signal and analyze the robustness of our network to the presence of device ID information in such portions. We would like the network to focus on nonlinear transmitter characteristics embedded in the packets rather than the device ID which can be easily spoofed. We expect these nonlinear features to be stable over time, as compared to the device ID which is localized in time. Here we focus on the ADS-B protocol and begin by describing its packet structure.

\subsection{ADS-B Packet Structure}

\begin{figure}[!t]
\centering
\vspace{6pt}
\includegraphics[width=0.98\columnwidth]{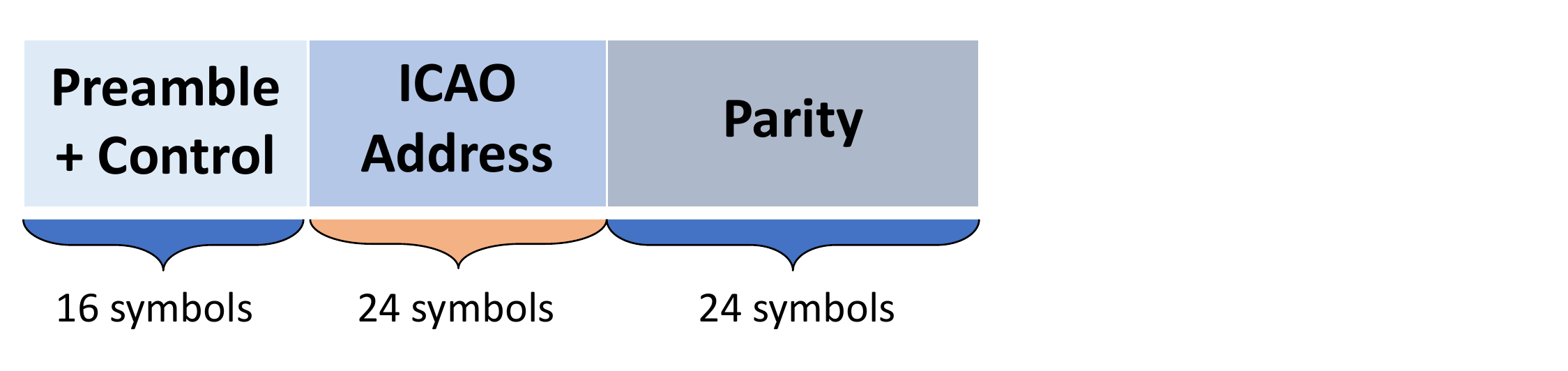}
\vspace{6pt}
\hspace{1.5pt}\includegraphics[width=0.98\columnwidth]{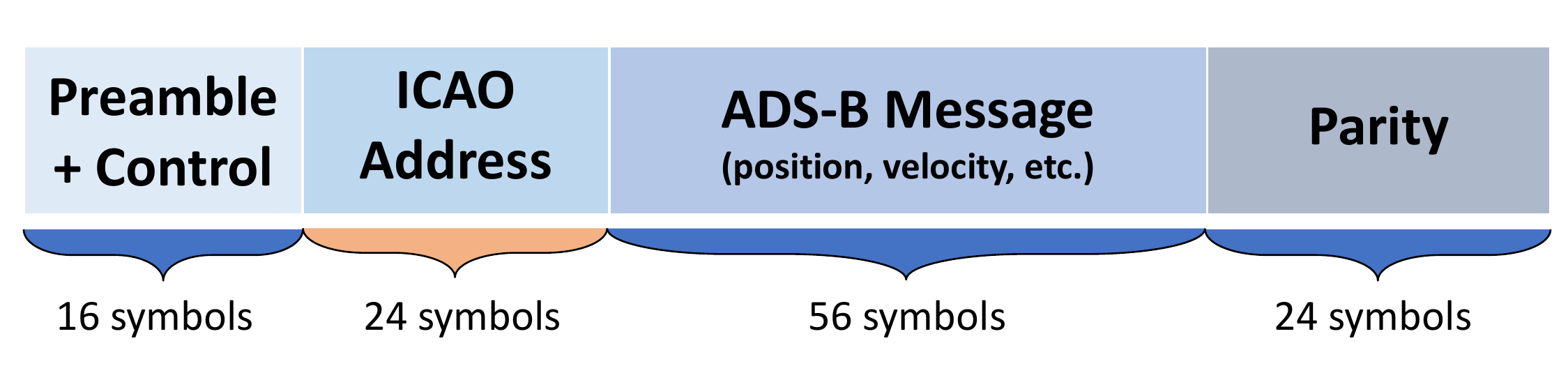}
\vspace{6pt}
\caption{Packet structure of ADS-B signals. Top: Mode S; bottom: Mode S Extended. The first 16 symbols of both packet types are device-independent, while the next 24 symbols are highly device-dependent.}
\label{fig:adsb_packet}
\vspace{-8pt}
\end{figure}

We consider two different types of ADS-B packets: Mode S and Mode S Extended, depicted in Fig. \ref{fig:adsb_packet}. For both packet types, the first 16 symbols consist of a preamble that is identical across devices, while symbols 17-40 contain the ICAO address which is unique to each device. The two modes have different packet lengths, with 64 symbols in Mode S and 120 symbols in Mode S Extended. For this reason, we prune Mode S Extended packets to 64 consecutive symbols, using an offset to determine the first selected symbol. 
We consider three different scenarios for the offset: an offset of zero, a randomly chosen offset and a fixed offset where we choose the last 64 symbols.

\subsection{Performance}

\begin{figure}[t]
\centering
\includegraphics[width=0.98\columnwidth]{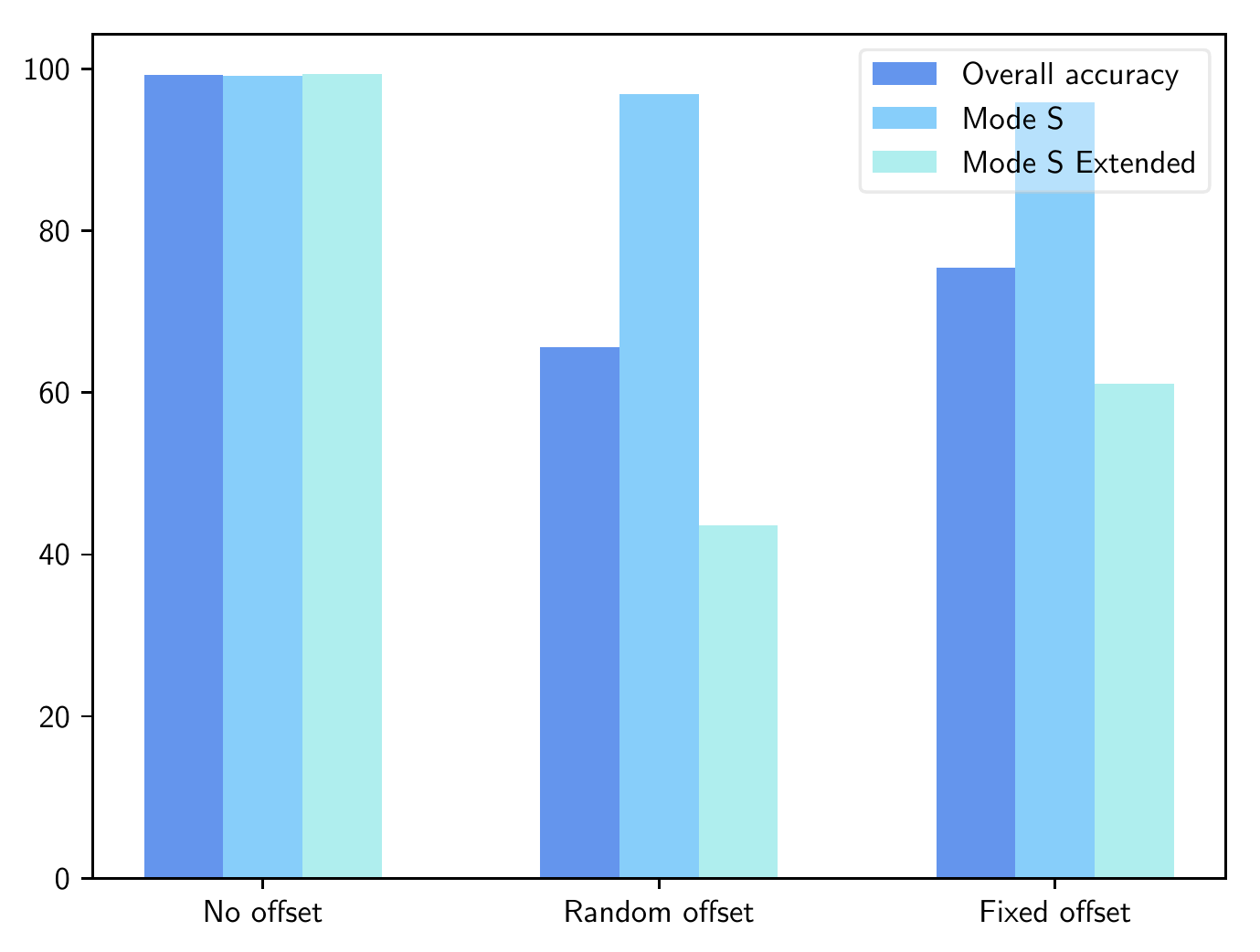}
\caption{Classification accuracies for ADS-B (100 devices) when using post-preamble data. 
Here we use architecture 100$\, C\,$100$\times$50 -- $|\cdot|^2$ -- 100$\, C\,$10$\times$2 -- Avg -- 100$\, D$.
For details on notation, see Section \ref{sec:arch_perf}.
 }
\label{fig:scenarios}
\vspace{-8pt}
\end{figure}

 Performance for each scenario is shown in Fig.\ \ref{fig:scenarios}. We report on accuracies for 100 devices, using 400 samples per device for training and testing. 
 We obtain a very high accuracy of 99.29\% when we do not use any offset, but this reduces to 65.64\% and 75.49\% in the scenarios with offsets. The picture becomes clearer when we examine the performance for Mode S and Mode S Extended: the two packet types have identical accuracies in the scenario without offset, but in the other scenarios, Mode S dominates performance. Such a temporal dependence indicates that the network is not learning the true nonlinearities, but rather focusing on device IDs from the payload for Mode S. It is easy to obtain 99\% accuracy by restricting attention to just the ICAO address (which can be easily spoofed), which is a clear indicator of ``cheating''. 

A natural approach to prevent such involuntary cheating might be to delete symbols 17-40 (which correspond to the ICAO address). However the presence of parity bits towards the end of the packet makes this approach insufficient. One can observe that a combination of parity and preamble sections can potentially reconstruct the ICAO address, and indeed in practice we obtain artificially high accuracies similar to the previous scenarios. In contrast, when we restrict attention to the preamble alone, performance decreases to 81.66\%, which is still much better than pure chance.
Another approach might be to set the kernel size of the first convolutional layer to 2 symbols, so as to prohibit the network from learning the ICAO address even if we allow access to the entire packet. This reduces accuracy to 97.28\%, but it is still much higher than when we use only the preamble. At first glance it may seem like small filter sizes at the first layer are sufficient to prevent cheating, but one just needs to look at the second layer to see that its filters actually extend over 6 symbols.

These experiments show that allowing networks to access ID information is unwise: networks ``cheat'' whenever given the chance and ignore transmitter-characteristic nonlinearities in favor of localized device information. We can mitigate this by allowing access to only the preamble, in which case we obtain the nonlinear fingerprints we are looking for.


\section{Robustness to Noise}

In this section, we investigate the impact of noise on classification accuracy. We first study the effect of real-world noise and then discuss noise augmentation strategies to enhance performance.

\subsection{Impact of naturally occurring noise}

We study the effect of different levels of noise in the training and test sets, using ADS-B data with 100 devices in each scenario.
When we cheat by using the ICAO address as described in the previous section, we obtain artificially high accuracies that are independent of the noise level, indicating that such a network can be easily spoofed even when the data is noisy.

When we use only the preamble, we observe a surprising trend (shown in Table \ref{table_adsb_experiments}): performance improves when the training data is \textit{noisier} than the test data. In contrast, when the training SNR is higher than the test SNR, we obtain high training accuracies but low test accuracies. While this result might seem initially counter-intuitive, it is a reasonable hypothesis that noise forces the network
to learn features that are more robust to perturbations.

\begin{table}[b]
\vspace{7pt}
\centering
\caption{Accuracy as a function of SNR for ADS-B (100 devices), using only the preamble. Here low SNR corresponds to $<$2 dB, medium SNR to 2-5 dB and high SNR to $>$5 dB.}
\label{table_adsb_experiments}
\begin{center}
\begin{small}
\begin{tabular}{llccc}
\toprule
Test SNR & Train SNR & Test accuracy & Train accuracy\\
\midrule
Low   & High   & 32.29 & 90.50 \\[2pt]
	 & Medium &  51.26 & 84.83 \\[2pt]
\cmidrule{2-4}
Medium & High   & 68.85 & 90.80 \\[2pt]
	 & Low    & 71.13 & 80.51  \\[2pt]
\cmidrule{2-4}
High  & Medium & 81.66 & 83.71 \\[2pt]
	 & Low    & 73.48 & 80.53  \\[2pt]
\bottomrule
\end{tabular}
\end{small}
\end{center}
\end{table}

\subsection{Noise augmentation}

\begin{table*}[t]
\centering
\caption{Effect of noise augmentation on ADS-B and outdoor WiFi fingerprinting. The ADS-B dataset corresponds to the first row of Table \ref{table_adsb_experiments} (with low test SNR, high train SNR). Noise injection improves ADS-B performance from \colorbox{LightGrey}{32.29\%} (which corresponds to train SNR$_{\text{aug}}$ = test SNR$_{\text{aug}}$ = $\infty$, i.e.\ no noise insertion) to \colorbox{LightCyan}{52.12\%} when train SNR$_{\text{aug}}$ = 10 dB and test SNR$_{\text{aug}}$ = 50 dB. Outdoor Wifi accuracy improves from \colorbox{LightGrey}{61.73\%} to \colorbox{LightCyan}{69.37\%}.
 }
\label{table_noise}
\vspace{3pt}
\begin{subtable}{0.49\linewidth}
	\caption{ADS-B (100 devices)}
	\label{table_noise_adsb}	
	\vspace{2pt}
	\centering
	\small
	\setlength\extrarowheight{4pt}
	\begin{tabular}{!{}c !{\color{black!70} \vrule} !{}c !{}c !{}c !{}c}%
		\arrayrulecolor{black!70}
		\color{black!30}
		\diagbox[width=2.2cm, height=1.3cm]{\color{black}\pbox{1.2cm}{Train SNR$_{\text{aug}}$\\[-5pt]}}{\color{black}\pbox{0.72cm}{Test \\SNR$_{\text{aug}}$} \hspace{2pt}}&
		20 dB & 50 dB & 100 dB & $\infty$ \\
		\arrayrulecolor{black!70}\hline
		10 dB  & 48.39 & 52.12\cellcolor{LightCyan} & 51.89 & 43.28 \\[2pt]
		15 dB & 52.12 & 50.75 & 51.98 & 40.63\\[2pt]
		20 dB & 35.25 & 47.63 & 45.44 & 15.29 \\[2pt]
		25 dB & 36.38 & 45.74 & 45.29 & 11.82 \\[2pt]
		$\infty$ & 25.67 & 33.55 & 34.77 & 32.29\cellcolor{LightGrey} \\[2pt]
	\end{tabular}
\end{subtable}
\hfill
\begin{subtable}{0.49\linewidth}
	\caption{WiFi (100 devices, outdoor environment)}
	\label{table_noise_wifi}
	\vspace{2pt}	
	\centering
	\small
	\setlength\extrarowheight{4pt}
	\begin{tabular}{!{}c !{\color{black!70} \vrule} !{}c !{}c !{}c !{}c}%
		\arrayrulecolor{black!70}
		\color{black!30}
		\diagbox[width=2.2cm, height=1.3cm]{\color{black}\pbox{1.2cm}{Train SNR$_{\text{aug}}$\\[-5pt]}}{\color{black}\pbox{0.72cm}{Test \\SNR$_{\text{aug}}$} \hspace{2pt}}&
		20 dB & 50 dB & 100 dB & $\infty$ \\
		\arrayrulecolor{black!70}\hline
		10 dB  & 61.65 & 62.21 & 61.90 & 3.04 \\[2pt]
		15 dB & 63.37 & 62.92 & 61.00 & 2.97 \\[2pt]
		20 dB & 69.37\cellcolor{LightCyan} & 69.06 & 67.83 & 2.09 \\[2pt]
		25 dB & 68.53 & 69.02 & 68.17 & 2.87 \\[2pt]
		$\infty$ & 29.90 & 31.45 & 30.93 & 61.73\cellcolor{LightGrey} \\[2pt]
	\end{tabular}
\end{subtable}
\end{table*}

We perform noise augmentation by inserting various levels of additional white Gaussian noise (AWGN) in the training and test sets, and report on accuracies as a function of injected noise levels in Table \ref{table_noise}. Here SNR$_{\text{aug}}$ denotes the signal to artificially injected noise ratio, so that SNR$_{\text{aug}}$ = $\infty$ corresponds to no noise injection. We consider two datasets: 100 ADS-B devices corresponding to the first row of Table \ref{table_adsb_experiments}, with 400 signals per device for training and testing; and 100 WiFi devices in an outdoor environment, with 800 signals per device for training and 200 for testing. 

Noise insertion yields significant performance benefits, with 19.83\% improvement for ADS-B and 7.64\% improvement for outdoor WiFi. We note, however, that it is important to add noise to both the training and test sets. Adding noise to only the training set can result in poor performance: for the WiFi data, at 20 dB train SNR$_{\text{aug}}$, test accuracy drops to 2.09\% (with 76.32\% training accuracy).


\section{Conclusions}

In this paper, we have demonstrated the efficacy of complex-valued CNNs for wireless fingerprinting. This technique does not rely on signal domain knowledge and,
as illustrated by our experiments with WiFi and ADS-B data, can be used across diverse wireless protocols. We show the vulnerability of the approach to ``cheating'' using transmitter ID
when using the entire message to extract the fingerprint. When using the preamble alone, reasonably high accuracies are obtained, and performance is significantly enhanced by noise augmentation.
Open issues worth investigating include (a) provably non-cheating, protocol-agnostic strategies that use the entire packet, (b) automatic extraction of the preamble given data corresponding to any protocol, (c) utilizing multiple packets for decisionmaking; and (d) developing detailed insight into the nature of the signatures
 extracted, and the impact of noise augmentation.


\section*{Acknowledgment}

This work was funded in part by DARPA under the AFRL contract number FA8750-18-C-0149, and by ARO under grant W911NF-19-1-0053. The views and conclusions contained herein are those of the authors and should not be interpreted as necessarily representing the official policies or endorsements, either expressed or implied, of DARPA or Air Force Research Laboratory or ARO or the U.S. Government. The authors gratefully acknowledge research discussions with collaborators at Teledyne Scientific, including Mark Peot, Laura Bradway, Karen Zachary and Michael Papazoglou.

\bibliographystyle{IEEEtran}
\bibliography{ms.bbl}

\end{document}